\documentclass[aps,pra,twocolumn,amssymb,amsmath,superscriptaddress]{revtex4-1}
\usepackage{graphicx,color,braket}
\usepackage[colorlinks = true]{hyperref}
\usepackage[T1]{fontenc}

\begin{document}
	
\title{Decoherence  can  help  quantum
  cryptographic  security}   \author{Vishal  Sharma}  \affiliation{IIT
  Jodhpur, Jodhpur, Rajasthan  India} \email{pg201383506@iitj.ac.in} \author{U.
  Shrikant}   \affiliation{Poornaprajna    Institute   of   Scientific
  Research, Bangalore -  562164, India} \email{shrik@poornaprajna.org}
\affiliation{Graduate  Studies,  Manipal  University,  Manipal-576104}
\author{R.     Srikanth}   \affiliation{Poornaprajna    Institute   of
  Scientific     Research,     Bangalore      -     562164,     India}
\email{srik@poornaprajna.org}       \author{Subhashish       Banerjee}
\affiliation{IIT          Jodhpur,           Rajasthan          India}
\email{subhashish@iitj.ac.in}

\begin{abstract}
In quantum key distribution, one conservatively assumes that the
eavesdropper Eve is restricted only by physical laws, whereas the
legitimate parties, namely the sender Alice and receiver Bob, are
subject to realistic constraints, such as noise due to
environment-induced decoherence.  In practice, Eve too may be bound by
the limits imposed by noise, which can give rise to the possibility
that decoherence works to the advantage of the legitimate parties.  A
particular scenario of this type is one where Eve can't replace the
noisy communication channel with an ideal one, but her eavesdropping
channel itself remains noiseless.  Here, we point out such a
situation, where the security of the Ping-Pong protocol (modified to a
key distribution scheme) against a noise-restricted adversary improves
under a non-unital noisy channel, but deteriorates under unital
channels.  This highlights the surprising fact that, contrary to the
conventional expectation, noise can be helpful to quantum information
processing. Furthermore, we point out that the measurement outcome
data in the context of the non-unital channel can't be simulated by
classical noise locally added by the legitimate users.  
\end{abstract}

	
\maketitle
 
\section{Introduction}
Cryptography  helps  secure  information  being  communicated  between
legitimate users  \cite{gisin2002quantum,srinatha2014quantum} across a
quantum       communication      channel       \cite{sharma2016effect,
  sharma2015controlled,   sharma2016comparative,  sharma2014analysis},
which    may   be    optical,    open    space   or    satellite-based
\cite{wang2014polarization,   sharma2017analysis}.     Since  the
seminal    BB84    quantum    key    distribution    (QKD)    proposal
\cite{bennett1984quantum}, the idea that unconditional security of the
distributed key  can be obtained  by using quantum resources  has been
extensively studied through more  detailed security analyses and newer
QKD protocols, among  them \cite{ekert1991quantum, bennett1992quantum,
  goldenberg1995quantum,    lo1999unconditional,   scarani2001quantum,
  lo2005efficient,       scarani2009security}.         See       Refs.
\cite{pathak2013elements} and  \cite{shenoy2017quantum} and references
therein.

A variant  of QKD is  one involving direct communication  avoiding the
step of key generation \cite{DL04}.  These protocols may be classified
as      QSDC      (quantum      secure      direct      communication)
\cite{bostrom2002deterministic, lucamarini2005secure,  SPS12} and DSQC
(deterministic   secure   quantum   communication)   protocols.    The
difference  is  that,  unlike  DSQC protocols,  QSDC  protocols  don't
require any  additional classical  communication, except  for checking
eavesdropping.  Other important cryptotasks under active investigation
include   quantum   coin   flipping   \cite{PCD+11},   quantum   money
\cite{amiri2017quantum}, quantum  private query \cite{wei2017generic},
quantum secure computation \cite{shi2016quantum}.

Environmental noise is ubiquitous in  the real world, and is generally
detrimental  to   quantum  communication  \cite{banerjee2008geometric,
  srikanth2008squeezed,  banerjee2007dynamics,  omkar2013dissipative}.
In quantum key distribution, it is  conservative to assume that all of
the  noise is  due  to an  eavesdropper Eve,  who  replaces the  noisy
(and/or lossy)  channel with  an ideal  one \cite{adhikari2015toward}.
Eve is assumed to  be as powerful as the laws  of physics would allow.
This determines the largest noise level that can be tolerated.

In  reality, we  may expect  that Eve,  too, to  be restricted  by the
noise. Alice, Bob and Eve may be  assumed to be aware of this.  As the
legitimate and eavesdropping channels are not identical, this scenario
of noise-restricted Eve gives rise to the interesting possibility that
noise may  be more  disadvantageous for  Eve than  for Alice  and Bob.
Here we shall  present a concrete instance of such  a situation.  This
can  be trivially  ensured by  making the  eavesdropping channel  more
noisy than Alice's and Bob'  communication channel. A more non-trivial
scenario is  one where  the noisy  channel acts  directly only  on the
communication channel  and not  on the  eavesdropping channel.  On the
other hand, Eve  is assumed to be unable to  replace the noisy channel
of  Alice  and  Bob  with  an  ideal  one.  

Our main  result is  the demonstration of  a quantum  key distribution
(QKD)  situation  where non-unital  noise  can  be beneficial  to  the
legitimate  participants  in  this  sense,  whereas  unital  noise  is
detrimental to  them.    This can potentially  form the  basis for
``trusted noise'', wherein Alice and  Bob add noise prior to classical
post-processing  to improve  the protocol's  security or  performance.
Interestingly, such  an application  of noise for  QKD has  been noted
earlier.   In particular,  in an  analysis of  various QKD  protocols,
\cite{renner2005information} shows  that they can be  made more robust
against channel noise by the addition of  noise by Alice or Bob to the
measurement    data    prior    to    key    reconciliation.     Refs.
\cite{pirandola2009direct, garcia-patron2009continuous} discuss adding
noise  to the  signal to  improve noise  tolerance in  the context  of
continuous-variable  QKD  over  Gaussian channels.   Interestingly,  a
somewhat  similar favorable  effect  of noise  on quantum  information
processing was noted in \cite{banerjee2008symmetry}. 

Secure  direct  communication  (SDC)  is a  stronger  form  of  secure
communication than key distribution  wherein message bits, rather than
random key  bits, are transmitted  from sender Alice to  receiver Bob.
Since  the proposal  of the  first  quantum SDC  protocol, namely  the
Ping-pong protocol \cite{bostrom2002deterministic},  a number of other
realizations of  this theme have been  proposed \cite{long2007quantum,
  wang2005quantum,        deng2004secure,        ting2005simultaneous,
  wang2005multi,  li2005quantum, jin2006three,  zhong2007improvement}.
The Ping-Pong protocol's  security, as well as  its modified versions,
have   been    extensively   studied   by   various    other   authors
\cite{wojcik2003eavesdropping,  han2014security, zawadzki2012security,
  zawadzki2012ping,       cai2004improving,      cai2004deterministic,
  cai2006eavesdropping,      zawadzki2016general,      li2012improved,
  zawadzki2012improving,      zhang2004improving,     wang2005quantum,
  vasiliu2011non, chamoli2009secure}.  A comprehensive review  of some
of the attacks  and protective measures against them  are discussed by
the authors of the Ping-pong protocol \cite{bostrom2008security}.

The original  Ping-pong protocol  is based on  two modes:  the message
mode during which a bit  is transmitted deterministically, and control
mode, to monitor eavesdropping. This  structure is necessitated by the
requirement for  the protocol to  perform as  a scheme for  SDC. Here,
however, we  will use a  simplified version of the  Ping-pong protocol
(though it  will still be called  as such), which is  suitable for key
distribution, but in general not for SDC. This is done by dropping the
control  mode, and  instead  using  a quantum  bit  error rate  (QBER)
analysis (which  involves sacrficing  some otherwise secret  bits) for
detecting eavesdropping.

For our purpose,  it will suffice to consider the  depolarizing and AD
(amplitude damping) channels, representative of unital channels (those
that map  the identity  operator to  itself) and  non-unital channels,
respectively. Furthermore,  the noise  acts only on  the communication
channel and not directly on the  eavesdropping channel, so that Eve is
affected  only  by  the  interaction  of her  probes  with  the  noisy
communication  channel,  rather  than   noise  acting  on  her  probes
directly. In  this scenario, the  semi-powerful Eve is able  to deploy
noiseless probes, but unable to replace Alice-Bob's noisy channel with
a noiseless one.

The remaining work is divided as follows.  In Section \ref{sec:PP}, we
briefly review  the Ping-Pong protocol  reformulated as a  QKD (rather
than    SDC)   scheme,    and    an   attacking    strategy   on    it
\cite{wojcik2003eavesdropping}.    In    Section   \ref{sec:env},   we
introduce  the  noise  scenario  used   in  this  work.   In  Sections
\ref{sec:AD}  and \ref{sec:depol},  we  study the  performance of  the
(modified)  Ping-Pong  protocol   in  the  presence  of   the  AD  and
depolarizing channels,  respectively, pointing out  the (surprisingly)
beneficial aspect of  the former.  The question  of the simulation
of the measurement outcome data under  a noisy channel by the resource
of  local  classical  noise  applied   by  the  legitimate  users,  is
considered in the conclusing Section \ref{sec:conclu}.
	    
\section{Eavesdropping on the Ping-Pong protocol \label{sec:PP}}

First, we  briefly describe the (modified)  Ping-Pong key distribution
protocol,  based on  the original  secure deterministic  communication
protocol  \cite{bostrom2002deterministic}.  In what  follows, we
use  the notation  where  $\ket{0}$ and  $\ket{1}$  represent the  two
polarization  states $H$  and $V$  of a  single photon,  respectively,
whilst $\ket{2}$  represents the  vacuum state.
\begin{enumerate}
\item Bob  transmits to Alice one  half (the ``travel qubit'')  of the
  Bell   state   $\ket{\psi^+}   =  \frac{1}{\sqrt{2}}   (\ket{01}   +
  \ket{10})$.
\item Alice encodes one bit of information
by applying operation $I$ (resp.,  Pauli $\sigma_Z$),
corresponding to the bit
value $a=0$ (resp.,  $a=1$).  
\item She retransmits the travel qubit back to Bob.
\item The two-qubit state now left with Bob is ideally in one the Bell
  states  $|\psi^{\pm}\rangle$,  which  is  determined  by  Bob  by  a
  Bell-state measurement.
\item For a  sufficiently large set of the (noisy)  shared bits, Alice
  announces the encoded bit on some of the transmissions. The fraction
  of  bits  where Alice's  and  Bob's  records differ  determines  the
  quantum bit  error rate (QBER).   If the  QBER is below  a threshold
  value, they proceed to distill a secret key. Else, they abort.
\end{enumerate}

W\"ojcik proposed an eavesdropping  strategy on the original ping-pong
protocol, which  is now adapted  for the modified  Ping-pong protocol.
The basic intuition of security in  the Ping-pong protocol is that the
travel qubit remains always in the maximally mixed state, irrespective
of  Alice's encoding.  The subtlety  of W\"ojcik's  attack is  that by
making the probe  interact before and after the encoding,  Eve is able
to extract some information about the encoding. A brief description
of the attack adapted to the above protocol is enumerated below.

\begin{enumerate}
\item  Eve  prepares  two  probes  $x$  and $y$  in  the  state  $|  2
  \rangle_{x}|0\rangle_y$, where $\ket{2}$ is  the vacuum state. Thus,
  the  combined initial  quantum  state  with Bob  and  Eve is  $|{\rm
    \psi^{initial}}\rangle   =   |\psi^{+}\rangle_{ht}   |2\rangle_{x}
  |0\rangle_{y}$.
\item In the onward leg, Eve  attacks the travel qubit by applying the
  operation $Q_{txy}  = SWAP_{tx}  CPBS_{txy} H_{y}$, with  CPBS being
  the controlled polarization beam splitter operation, given by:
\begin{equation}
\left. \begin{array}{c}
\ket{020} \\
\ket{021} \\ \ket{120} \\ \ket{121}
\end{array} \right\}
~\stackrel{CPBS}{\longrightarrow}~
\left\{ \begin{array}{c}
\ket{002} \\
\ket{021} \\ \ket{120} \\ \ket{112}
\end{array} \right.
\end{equation}
\item After  Alice has  encoded her  bit on the  travel qubit  and she
  returns it, Eve  applies the operation $Q_{txy}^{-1}$  on the travel
  qubit and forwards it to Bob.
\end{enumerate}

Eve then obtains some information  about Alice's encoding by measuring
her probes. To  see how the attack  works, we note that  after Bob has
received  back the  attacked  travel  qubit, the  final  state of  the
Alice-Bob-Eve system is
\begin{equation}
|\psi^a\rangle_{htxy} = \frac{1}{\sqrt{2}}(\ket{012a} + \ket{1020}).
\label{eq:Eve}
\end{equation}

From  this,  one  finds  that  the  only  non-vanishing  probabilities
$P_{AEB}$ are
\begin{align}
P_{000}&=\frac{1}{2} \nonumber \\
P_{100}&= P_{101}=P_{110}=P_{111}=\frac{1}{8}.
\label{eq:probs}
\end{align}
This   corresponds  to   a   QBER  of   $\sum_e  (p_{0e1}+p_{1e0})   =
\frac{1}{4}$.   Using these,  one may  compute the  mutual information
between Alice and Bob, $I_{AB} \equiv H(A) - H(A|B)$, where $H(A)$ and
$H(A|B)$  are   the  classical   (Shannon)  entropy   associated  with
probability  distribution  $P(a)$   and  the  conditional  probability
distribution $P(a|b)$ \cite{nielsen2010quantum}.  This is a measure of
entropic correlation  between Alice  and Bob.  Similarly,  one defines
the mutual information  between Alice and Eve,  given by $I_{AE}\equiv
H(A)-H(A|E)$.    From   (\ref{eq:probs}),    one   then   finds   that
\cite{wojcik2003eavesdropping}
\begin{equation}
I_{AB}  = I_{AE} = \frac{3}{4}\log_{2}\frac{4}{3} \approx 0.311.
\label{eq:entrop}
\end{equation}
Thus, the  attack makes  the protocol  insecure, since  security (with
one-way communication) requires that $I_{AB}>I_{AE}$. 

This  attack   is  not   symmetric  between   $a=0$  and   $a=1$,  and
\cite{wojcik2003eavesdropping}  proposes  another,  symmetric  attack.
Ref. \cite{bostrom2008security} discusses a number of other attacks on
the Ping-pong  protocol, showing it  to be effectively  robust against
them.  Thus, while the attack described is not known to be optimal, it
represents  a powerful  and well-studied  attack, and  its performance
under decoherence is  likely to carry general implications  of a wider
nature, in particular the occurence of the ``trusted noise'' scenario.
Therefore, our present work is focused on studying this aspect of it.

Furthermore,  it is  generally difficult  to prove  the security  of a
given  QKD protocol  against  the most  general (collective)  attacks,
though, specific protocols can be  proposed where such security can be
proven.  Under  the circumstances, a  reasonable approach is  to prove
security  against   a  non-general,  but  sufficiently   powerful  and
sophisticated attack, which is the case here.

\section{Quantum communication under a noisy environment\label{sec:env}}

\begin{figure}
             \includegraphics[width=7cm]{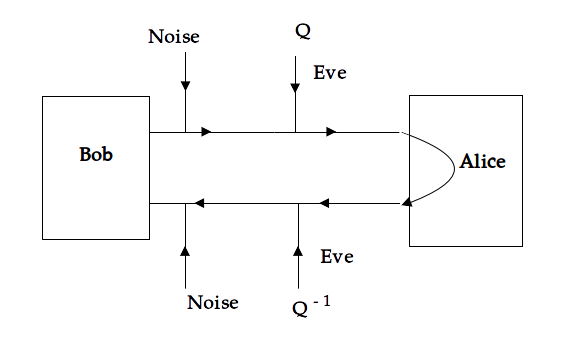}
\caption{Scenario  of noise  and  attack  as used  in  this work:  Bob
  transmits to Alice one half of  a Bell state, on which Alice encodes
  her bit by  applying either the operation $I$  or $\sigma_Z$, before
  returning it  to Bob.  The  action of  noise is idealized  as acting
  before  Eve's action  $Q$ in  the onward  leg and  after her  action
  $Q^{-1}$ in the return leg.}
          \label{fig:attacknoise}
          \end{figure}

The action of noise manifesting as a completely positive (CP) map on a
system's density operator, can be given a Kraus representation:
	    \begin{equation}
	    \phi(\rho) =\sum_{i} A_{i}\rho A^\dag_{i},
\label{eq:the 51-qubit state}
	    \end{equation}
where  the  $A_{i}'s$  must  conform to  the  completeness  constraint
$\sum_{i}A^\dag_{i}A_{i} = I$.  In this work, we choose the simplified
noise scenario depicted in Figure \ref{fig:attacknoise}. In the onward
leg,  the noise  first acts  on the  travel qubit,  followed by  Eve's
attack $Q$ on  this qubit, and then by Alice's  encoding operation. In
the return leg,  this sequence is time-reversed, so  that Eve's second
attack $Q^{-1}$ is followed by the noise, before receipt of the travel
qubit and decoding of the two-qubit state by Bob.

In  a noisy  channel, suppose  bits 0  and 1  correspond to noisy
states $\rho^{a=0}$  and $\rho^{a=1}$. Then, the mutual  information between
the Alice and Bob is upper-bounded by the Holevo bound:
\begin{equation}
\chi = S\bigg(\frac{\rho^{a=0}_{\rm ht}+\rho^{a=1}_{\rm ht}}{2}\bigg)
- \frac{1}{2}\bigg[S\bigg(\rho^{a=0}_{\rm ht}\bigg)+
S\bigg(\rho^{a=1}_{\rm ht}\bigg)\bigg],
\label{eq:holevo}
\end{equation}
where  $S(\rho) \equiv  -\textrm{Tr}[\rho\log(\rho)]$ denotes  the von
Neumann entropy.

We  next consider  noisy conditions  with  Eve's above  attack on  the
Ping-Pong  QKD  protocol,  with  the travel  qubit  subjected  to  the
amplitude damping (AD) \cite{deleter} and depolarizing channels.

\subsection{Amplitude-Damping Noise\label{sec:AD}}

The Kraus operators for AD channel are \cite{srikanth2008squeezed}:
\begin{equation}
		E_{0}^{A}=\left[\begin{array}{cc}
		1 & 0\\
		0 & \sqrt{1-p}
\end{array}\right];\,\, E_{1}^{A}=\left[\begin{array}{cc}
		0 & \sqrt{p}\\
		0 & 0
		\end{array}\right],\label{eq:Krauss-amplitude-damping}
		\end{equation}
where  $p$  is the  noise parameter,  sometimes
called the decoherence rate, and $ 0 \le p \le 1 $.
	
The first attack  of \cite{wojcik2003eavesdropping} (during the onward
leg) makes the channel  lossy and
involves creating  the  vacuum  state   of  the  travel   photon. This
necessitates      extending      the       qubit      noise      model
(\ref{eq:Krauss-amplitude-damping}) to that of a qutrit.  There is no unique way to do this.  

We use the extension represented by the following Kraus operators:
\begin{equation}
E_{0}^{A}=\left[\begin{array}{ccc}
1 & 0 & 0 \\
0 & \sqrt{1-p} & 0 \\
0 & 0 & 1 
\end{array}\right];\,\, 
E_{1}^{A}=\left[\begin{array}{ccc}
0 & \sqrt{p} & 0\\
0 & 0 & 0 \\
0 & 0 & 0
\end{array}\right],\label{eq:Krauss-amplitude-damping-1}
\end{equation}
which       essentially       implements      the       AD       noise
Eq.  (\ref{eq:Krauss-amplitude-damping}) on  the polarization  Hilbert
space and does nothing to the vacuum state. Here  the vacuum state is taken to be the third  dimension,
denoted  $|2\rangle$. 

\begin{figure}
\centering
\includegraphics[width=0.48\textwidth]{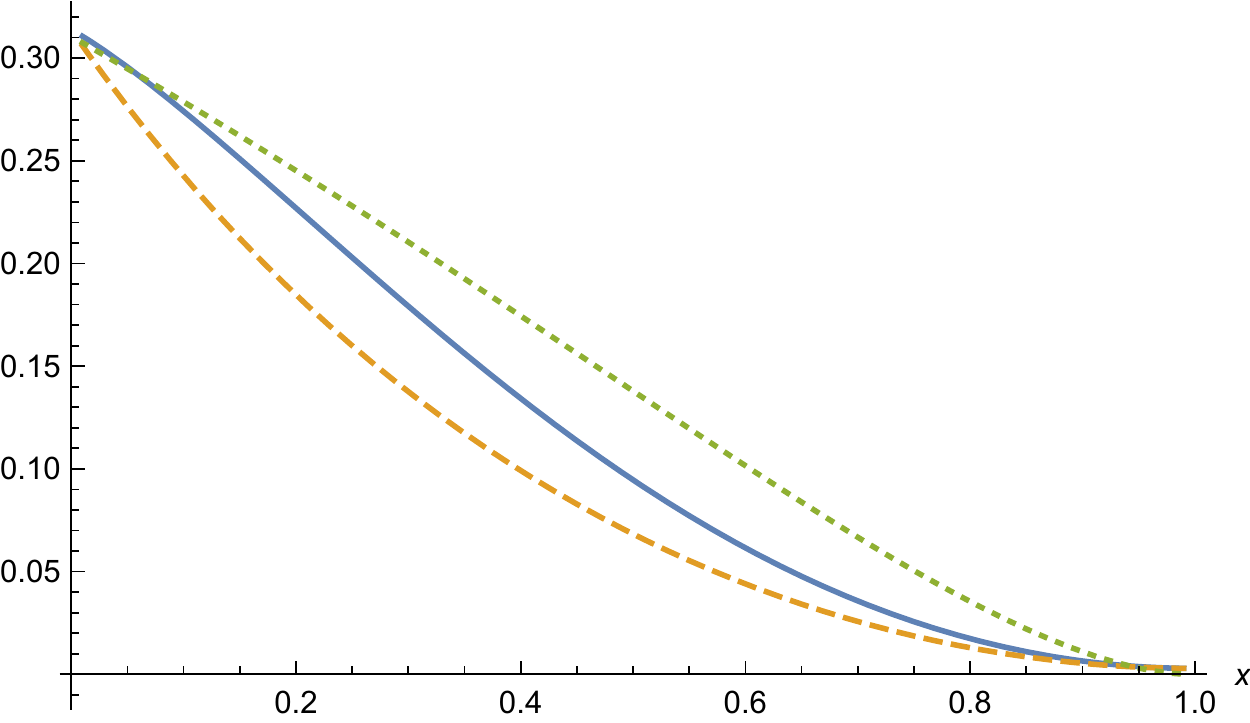}
\caption{(Color online) Performance of the modified Ping-pong protocol
  under AD  noise: The  bold (blue), dashed  (red) and  dotted (green)
  plots  represents  $I_{AB}$,  $I_{AE}$  and  the  Holevo  bound  for
  Alice-Bob. That $I_{AB} > I_{AE}$ for $0<p \le 1$ implies that noise
  is beneficial to the legitimate  users.  In the noiseless limit, the
  Holevo bound coincides with  $I_{AB}$, implying that the measurement
  strategy is optimal.}
\label{fig:AD}
\end{figure}

When  the photon  returns  back  to Bob,  the state of the system ${hty}$ for either encoding '$a$' can be shown to
have support  of dimensionality 4,  spanned by the  states $\ket{010},
\ket{100},  \ket{011}$ and  $\ket{000}$,  with the  state  of the  $x$
particle being $\ket{2}$, as in  the noiseless attack case.  The final
states with Bob-Eve for the encodings $a=0$ and $a=1$ are:
\begin{eqnarray}
\rho^{a=0}_{hty} =\frac{1}{2} \left(
\begin{array}{cccc}
(1-p)^2 & 1-p & 0 & 0 \\
1-p & 1 & 0 & 0 \\
0 & 0 & p (2-p) & 0 \\
0 & 0 & 0 & 0 \\ 
\end{array}    
\right) ; \nonumber  \\
\rho^{a=1}_{hty} = \frac{1}{2} \left(
\begin{array}{cccc}
0 & 0 & 0 & 0 \\
0 & 1 & 0 & 1-p \\
0 & 0 & p (2-p) & 0 \\
0 & 1-p & 0 & (1-p)^2 \\ 
\end{array}
\right).
\label{eq:AD}
\end{eqnarray}
From Eq.   (\ref{eq:AD}), we obtain the  following joint probabilities
$p_{AEB}$, in place of Eq. (\ref{eq:probs}):
	\begin{align}
		P_{000} &= \frac{1}{8} (2-p)^2 \nonumber \\
		P_{001} &=  \frac{p^2}{8}\nonumber  \\
		P_{002} &= P_{003} = P_{102} =P_{103}=   \frac{1}{8} (2-p) p \nonumber \\
		P_{110} &= P_{111} =  \frac{1}{8} (1-p)^2 \nonumber \\
		P_{010} & =P_{011} = P_{012} = P_{013} = 0 \nonumber \\ 
		P_{100} &= P_{101} =\frac{1}{8},
		\label{eq:ADjp2}
		\end{align}		
with all  other joint probability  terms vanishing.  Note that  in the
presence of AD noise, Bob  will also obtain outcomes $|\phi^\pm\rangle
=   \frac{1}{\sqrt{2}}(\ket{00} \pm \ket{11})$    in   his    Bell   state
measurement,  which corresponds  to  the  outcome symbols  2  and 3  in
Eq. (\ref{eq:ADjp2}).

From the above probabilities  $P_{AEB}$, one derives  the mutual
information between Alice  and Bob and that between Alice  and Eve, to
be
\begin{widetext}
\begin{align}
I(A:B) &= \frac{1}{8}\bigg[
p \left(p \log \left(\frac{p^2}{2 p^2-2 p+2}\right)+p \log \left(\frac{8 (p-2)^2}{(p-3) p+3}\right)+(p-2) \log \left(\frac{(p-2) p+2}{(p-1) p+1}\right) \nonumber \right. \\ &+
\left.(p-2) \log \left(\frac{p-1}{(p-3) p+3}+1\right)\right)-2 p (p+2) \log (2)-4 (p-1) \log \left(\frac{(p-2)^2}{2 ((p-3) p+3)}\right)
\nonumber \\ &+2 \log \left(\frac{(p-2) p+2}{(p-1) p+1}\right)+2 \log \left(\frac{p-1}{(p-3) p+3}+1\right)+4
 \bigg],
\label{eq:ADIAB}
\end{align}
\end{widetext}
and 
\begin{align}
I(A:E) &= \frac{1}{8} \bigg( 6 + 2 \log \left(\frac{1}{-p^2+2 p+3}\right)
\nonumber \\ &+(1-(p-2) p) \log \left(\frac{(p-2) p-1}{(p-3) (p+1)}\right) \bigg), 
\end{align}
respectively. These two quantities are depicted as a function of noise
$p$ in  Figure \ref{fig:AD}.   This shows that  under the  AD channel,
there  is a  positive key  rate $\kappa  \equiv I_{AB}  - I_{AE}$  for
finite noise.  It  is as if the symmetry existing  between Bob and Eve
in  terms of  information  gained,  is broken  by  the  noise, to  the
advantage of Alice and Bob.  This  is a surprising result, and implies
that Alice  and Bob will  find this type  of noise beneficial  in this
eavesdropping scenario.

If Alice and Bob are employing  the original Ping-pong strategy and the
eavesdropper     is    known     to    employ     the  above  attack, then in  the noise range $0 <  p < 1$,
Alice and  Bob know that they  can extract a finite  secret key, after
suitable  privacy  amplification.

From Eq. (\ref{eq:AD}) one obtains the reduced density operators for
the particles $ht$:
\begin{eqnarray}
\rho^{a=0}_{ht} &=\frac{1}{2} \left(
\begin{array}{ccc}
(1-p)^2 & 1-p & 0 \\
1-p & 1 & 0 \\
0 & 0 & p (2-p) \\ 
\end{array}    
\right) \, \, ; \nonumber  \\
\rho^{a=1}_{ht} &=\frac{1}{2} \left(
\begin{array}{ccc}
(1-p)^2 & 0 & 0 \\
0 & 1 & 0 \\
0 & 0 & p (2-p) \\
\end{array}
\right).
\label{eq:ADht}
\end{eqnarray}
in  the   basis  $\{\ket{01},  \ket{10},  \ket{00}\}$.    

The maximum information Bob can  receive is upper-bounded by the Holevo
quantity  (\ref{eq:holevo}). To  obtain this,  we note that the eigenvalues
$\lambda^0_j,  \lambda^1_j$  and   $\lambda^{01}_j$  for  the  density
operators $\rho^{a=0}_{ht}, \rho^{a=1}_{ht}$  and their equal average, are:
\begin{align}
\lambda^0_j &= \left\{0,-\frac{1}{2} (p-2) p,\frac{1}{2} ((p-2) p+2)\right\}
\nonumber\\
\lambda^1_j &= \left\{\frac{1}{2},\frac{1}{2} (p-1)^2,-\frac{1}{2} (p-2) p\right\}\nonumber\\
\lambda^{01}_j &= \left\{\frac{(2-p)}{2} p,\frac{1}{4} \left((p-2) p
\pm \sqrt{(p-2) p (p-1)^2+1}+2\right)\right\}
\label{eq:ADev}
\end{align}
The Holevo bound (\ref{eq:holevo})  is thus given by:
\begin{equation}
\chi_{AD} = h\left[\lambda^{01}_j\right] - 
\frac{1}{2}\left(h\left[\lambda^0_j\right] + h\left[\lambda^1_j\right]
\right),
\label{eq:ADholevo}
\end{equation}
where $h\left[\lambda^\alpha_j\right] = -\sum_{j=0}^2 \lambda^\alpha_j
\log_2\left(\lambda^\alpha_j\right)$.  The   quantity  $\chi_{AD}$  is
plotted in Figure \ref{fig:AD}.


That the  Holevo bound exceeding  $I_{AB}$ here suggests
that Bob's Bell state measurement strategy, although
guaranteeing a  positive key  rate, is sub-optimal.  Note that  it is
indeed optimal in the noiseless case. \\

\subsection{Depolarizing noise\label{sec:depol}}

Consider the travel qubit subjected  to depolarizing noise. This noise
is   characterized  by   the   transformation  $\rho   \longrightarrow
p\frac{\mathbb{I}}{2}     +      (1-p)\rho     $     \cite{nielsen2000quantum,
  omkar2013dissipative}, for which the Kraus operators are:
\begin{align}
 D_{0}&=\sqrt{1-p}\left( \begin{array}{cc} 1 & 0 \\   
			          0 & 1 \end{array} \right),
D_{1} = \sqrt{\frac{p}{3}}\left( \begin{array}{cc} 0 & 1 \\
			           1 & 0 \end{array} \right);
\nonumber\\
D_{2}&=\sqrt{\frac{p}{3}}\left( \begin{array}{cc} 0 & -i \\ 
			          i & 0 \end{array} \right),
D_{3}=\sqrt{\frac{p}{3}}\left( \begin{array}{cc} 1 & 0 \\  
			            0 & -1 \end{array} \right),
\label{eq:2depolarizing}
\end{align}
where  $p  = (1-\exp^{-\frac{\tau  t}{2}})$,  $\tau$  being the  decay
factor. Here  we shall use the  extension of Eq.
 (\ref{eq:2depolarizing}) given by:
\begin{align}
 D_{0}&=\sqrt{1-p}\left( \begin{array}{ccc} 1 & 0 & 0 \\   
			          0 & 1 & 0 \\ 0 & 0 & 1\end{array} \right),
D_{1} = \sqrt{\frac{p}{3}}\left( \begin{array}{ccc} 0 & 1 & 0\\
			           1 & 0 & 0 \\ 0&0&1\end{array} \right);
\nonumber\\
D_{2}&=\sqrt{\frac{p}{3}}\left( \begin{array}{ccc} 0 & -i  & 0\\ 
			          i & 0 & 0 \\ 0&0&1\end{array} \right),
D_{3}=\sqrt{\frac{p}{3}}\left( \begin{array}{ccc} 1 & 0 &0\\  
			            0 & -1 &0\\0&0&1\end{array} \right),
\label{eq:3depolarizing}
\end{align}
which essentially implements a  depolarizing noise on the polarization
Hilbert space and does nothing to the vacuum state.
	
\begin{figure}[h]
\centering
\includegraphics[width=0.48\textwidth]{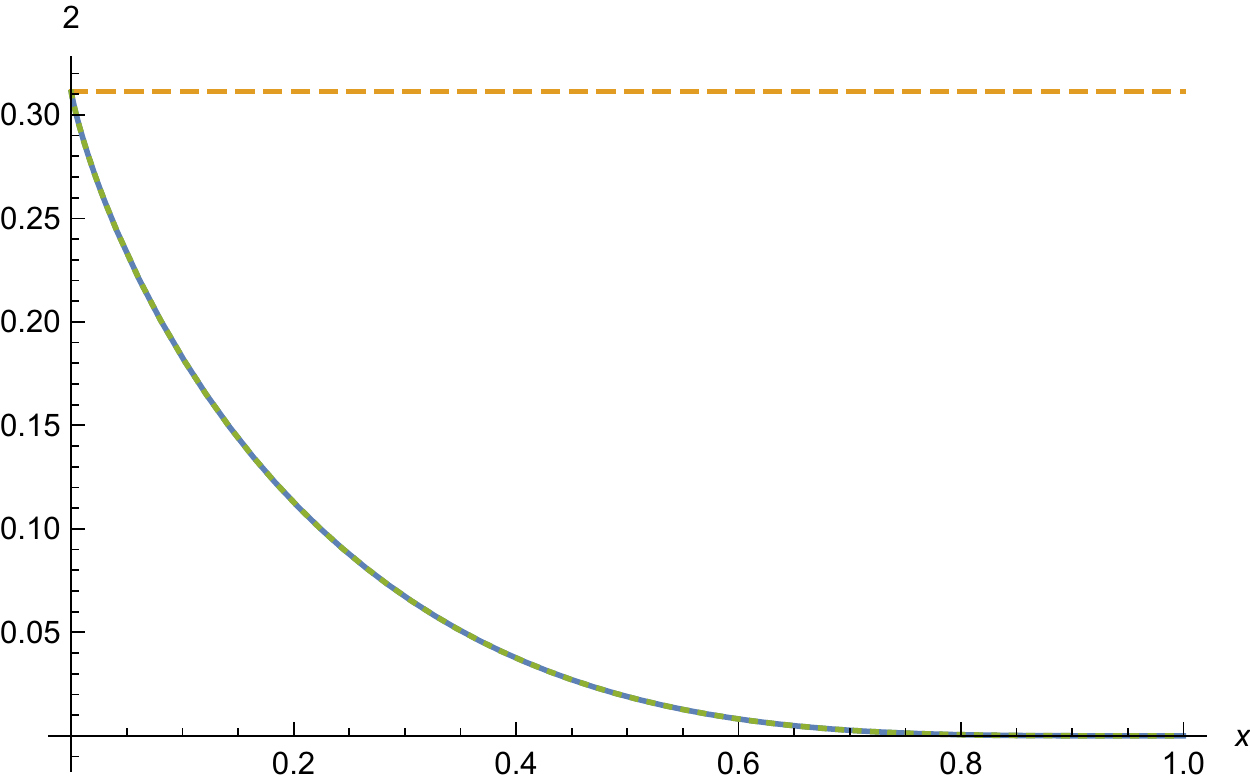}
\caption{(Color  online) Performance  of  the  Ping-pong QKD  protocol
  under depolarizing  noise: The bold  (black) and dashed  (red) plots
  represent $I_{AB}$ and $I_{AE}$ with  the Holevo bound for Alice-Bob
  coinciding with  $I_{AB}$.  As  a function  of noise  parameter $p$,
  $I_{AE}$ remains constant  at the noiseless value  of 0.311, because
  Eve's attack strategy is indifferent to unital noise.  That $I_{AB}$
  equals the Holevo bound implies that Bob's Bell state measurement in
  the modified Ping-pong protocol is already optimal.}
\label{fig:dpol}
\end{figure}

When the  photon returns back to Bob, as  per the scenario  of Figure
\ref{fig:attacknoise},  with  the  noise  given  by  the  depolarizing
channel, the state  of the system ${hty}$ for either  encoding $a$ can
be shown to have support of dimensionality 8, spanned by the states
$\ket{jkl}$, with $j,k,l \in \{0,1\}$, and the state of the $x$ particle
being $\ket{2}$  as in  the noiseless attack  case.  The  final states
with Bob-Eve for the encodings $a=0$ and $a=1$ are:
\begin{widetext}
	\begin{align}
		\rho^0 &= \frac{1}{2}\left(
		\begin{array}{cccccccc}
			\frac{p (4-p)}{4}  & 0 & 0 & 0 & 0 & 0 & 0 & 0 \\
			0 & 0 & 0 & 0 & 0 & 0 & 0 & 0 \\
			0 & 0 & \frac{(p-2)^2 }{4} & 0 & (p-1)^2 & 0 & 0 & 0 \\
			0 & 0 & 0 & 0 & 0 & 0 & 0 & 0 \\
			0 & 0 & (p-1)^2 & 0 & \frac{(p-2)^2}{4}  & 0 & 0 & 0 \\
			0 & 0 & 0 & 0 & 0 & 0 & 0 & 0 \\
			0 & 0 & 0 & 0 & 0 & 0 & \frac{p(4-p) }{4} & 0 \\
			0 & 0 & 0 & 0 & 0 & 0 & 0 & 0 \\
		\end{array}
		\right) \, \, ; \nonumber \\
		\rho^1 &= \frac{1}{2} \left(
		\begin{array}{cccccccc}
			\frac{p(2-p)}{4}   & 0 & 0 & 0 & 0 & 0 & 0 & 0 \\
			0 & \frac{p (2-p)}{4} & 0 & 0 & 0 & 0 & 0 & 0 \\
			0 & 0 & \frac{p^2}{4} & 0 & 0 & 0 & 0 & 0 \\
			0 & 0 & 0 & \frac{(p-2)^2}{4}  & (p-1)^2 & 0 & 0 & 0 \\
			0 & 0 & 0 & (p-1)^2 & \frac{(p-2)^2}{4} & 0 & 0 & 0 \\
			0 & 0 & 0 & 0 & 0 & \frac{p^2}{4} & 0 & 0 \\
			0 & 0 & 0 & 0 & 0 & 0 & \frac{p(2-p)}{4}  & 0 \\
			0 & 0 & 0 & 0 & 0 & 0 & 0 & \frac{p(2-p) }{4} \\
		\end{array}
		\right) .
\label{eq:depol}
	\end{align}
\end{widetext}
From Eq.   (\ref{eq:depol}), we obtain the  following joint probabilities
$P_{AEB}$, in place of Eq. (\ref{eq:probs}):
\begin{align}
		P_{000} &= \frac{1}{2} +  \frac{3p}{8} (p-2) \nonumber\\
		P_{001}  &= P_{002} = P_{003} = \frac{p}{8} (2-p)  \nonumber\\
		P_{010} &=P_{011} =P_{012} = P_{013}  = 0 \nonumber\\
		P_{100} &= P_{101} =P_{110} = P_{111} 	 =  \frac{1}{8} + \frac{p}{16} (p-2) \nonumber\\
		P_{102} &=P_{103} = P_{112} = P_{113} = \frac{p}{16} (2-p)
\label{eq:probdepol} 
\end{align}
with all other  joint probability terms vanishing.  As  with AD noise,
here  again   Bob  will  also  obtain   outcomes  $|\phi^\pm\rangle  =
\frac{1}{\sqrt{2}}(\ket{00} \pm \ket{11})$ in his  Bell state measurement,
which   correspond   to    the   outcome   symbols   2    and   3   in
Eq. (\ref{eq:probdepol}).

From  the  above  probabilities  $P_{AEB}$ ,  one  finds  the  mutual
information between Alice and Bob to be
\begin{align}
I_{AB}(p) &= \frac{1}{36} \bigg[
9 \log\left(\frac{4}{9} p (2 p-3)+1\right) \nonumber \\ &+ 
8 p (2 p-3) \coth ^{-1}\left(\frac{9}{(3-4 p)^2}\right) \nonumber\\
&+ 6 \left(4 p^2-6 p+3\right) \log\left(\frac{3}{4}-\frac{9}{64 p (2 p-3)+108}\right) \nonumber \\
&+(4p(2p-3)+9)\log\left(\frac{36}{16p(2 p-3)+27}+4\right)\bigg].
\end{align}

On the other hand, it follows from Eq. (\ref{eq:probdepol}) that
\begin{align}
P_{AE=00} &=\frac{1}{2}; P_{AE=01} =0 \nonumber\\
P_{AE=10} &= P_{AE=11} = \frac{1}{4},
\label{eq:depolEve}
\end{align}
i.e., $P_{AE}$  is independent of the  noise parameter.  Consequently,
$I_{AE}(p)$      is     just      the      noiseless     value      of
$\frac{1}{8}\log\left(\frac{64}{27}\right)$. 

Figure  (\ref{fig:dpol}) shows  that under  the depolarizing  channel,
there is  no positive  key rate  $\kappa \equiv  I_{AB} -  I_{AE}$ for
finite  noise, essentially  because $I_{AE}$  remains constant,  being
unaffected by  the depolarizing noise  ( as explained  above), whereas
$I_{AB}$  drops with  the noise  level.  Therefore,  this channel,  in
contrast to  the AD channel, offers  no advantage to Alice  and Bob in
our scenario.

 A similar disadvantageous behavior  holds for dephasing and other
unital noisy channels,  which may be understood  generally as follows.
In our  scenario, the noise  acts \textit{before} the first  attack by
Eve  (see Figure  \ref{fig:attacknoise}), and  the second  instance of
noise (in the backward trip of the particle) acts \textit{after} Eve's
second attack.

Therefore,  the  second  instance  of noise  doesn't  affect  $I_{AE}$
(though, in general,  it will affect $I_{AB}$). As to  the onward trip
of the particle, the  travel qubit, as seen by Eve,  is initially in a
maximally mixed state $\frac{I}{2}$.   Depolarizing noise or any other
unital channel $\mathcal{C}_U$ is characterized by the property
\begin{equation}
\mathcal{C}_U: \frac{I}{2} \mapsto \frac{I}{2},
\label{eq:mapsto}
\end{equation}
ie., it maps  the state $\frac{I}{2}$ to itself.  Thus,  this state of
the travel qubit remains unaffected,  and hence Eve's correlation with
Alice is indifferent to the noise.

It is  worth noting  here that  if the unital  noise acts  after Eve's
first    intervention     (rather    than    before,     see    Figure
\ref{fig:attacknoise}), then $I_{AE}$ is  not expected to be invariant
under  the noise,  since Eve's  action can  deviate the  state of  the
particle from $\frac{I}{2}$.

From Eq. (\ref{eq:depol}) one obtains the reduced density operators for
the state of particles $ht$
{\small
	\begin{equation}
	\begin{array}{lcl}
	\rho^{a=0}_{ht}=
	\frac{1}{4}\left(\begin{array}{cccc}
	(2-q)q & 0 & 0 & 0 \\
	0 & ((q-2)q+2) & 2(q-1)^2 & 0 \\
	0 & 2(q-1)^2 & (q-2)q+2 & 0 \\
	0 & 0 & 0 & (2-q)q \\
	\end{array}\right)\, \, ; \nonumber \\ 
	\rho^{a=1}_{ht}=
	\frac{1}{4}\left( \begin{array}{cccc}
	(2-q)q  & 0 & 0 & 0 \\
	0 & (q-2)q+2 & 0 & 0 \\
	0 & 0 & (q-2)q+2 & 0 \\
	0 & 0 & 0 & (2-q)q  \\
	\end{array}
	\right).
	\end{array}
	\label{eq:depolht}
	\end{equation}
}

As with Eq. (\ref{eq:ADev}), the maximum information Bob can receive is
upper-bounded by the Holevo  quantity (\ref{eq:holevo}).  To derive this,
we   obtain    the   eigenvalues   $\lambda^0_j,    \lambda^1_j$   and
$\lambda^{01}_j$   for   the   density   operators   $\rho^{a=0}_{ht},
\rho^{a=1}_{ht}$ and their equal average,  which are found to
be:
\begin{equation}
\begin{array}{lcl}
\lambda^0_j = 
\frac{1}{4} \left\{(2-p)p, (2-p)p,(2-p)p, 3 (p-2) p+4\right\},
\nonumber\\
\lambda^1_j= \frac{1}{4} \left\{(2-p)p, (2-p)p,(p-2) p+2, (p-2) p+2 \right\},\nonumber\\
\lambda^{01}_j= \frac{1}{4}\left\{1,(2-p) p, (2-p)p,2 (p-2) p+3 \right\}.
\end{array}
\label{eq:depolev}
\end{equation}	
Using  this,  the  Holevo  bound $\chi_{DP}$  under  the  depolarizing
channel can be found in a manner similar to Eq. (\ref{eq:ADholevo}). Interestingly $ \chi_{DP} $ is found to coincide with $ I_{AB} $.
This coincidence suggests that the Bell state measurement strategy by  Bob is indeed
optimal, unlike in the case of the AD channel.
	
\section{Conclusion and discussions \label{sec:conclu}}

It  is  generally  accepted  that  noise  is  detrimental  to  quantum
information processing,  in particular quantum cryptography.   Here we
identify,  counter  to  this  expectation,  a  scenario  of  ``trusted
noise'',  where  noise  can  play  a helpful  role.   In  quantum  key
distribution,  proofs  of  unconditional   security  assume  that  the
eavesdropper Eve is restricted only by physical laws, and that all the
noise is  due to her attack.   We consider a more  realistic scenario,
where  Eve   too  is  bound  by   limits  imposed  by  noise   due  to
environment-induced decoherence.   We show  how this  can work  to the
advantage of  legitimate parties, when noise  affects the eavesdropper
more  than the  legitimate  parties.   Now, an  easy  version of  this
scenario would have been one, where noise universally affects not just
the legitimate parties, but also Eve. Therefore, the nontrivial aspect
is that the  noise only affects the communication channel  and not the
eavesdropping channel directly.  Eve's  limitation is her inability to
replace the  noisy communication channel  between Alice and Bob  by an
noiseless  one.   In the  particular  situation  considered here,  the
security of  the Ping-Pong  protocol (modified  to a  key distribution
scheme) against a noise-restricted adversary is shown to improve under
a non-unital decoherence, but to deteriorate under unital decoherence.

In   light    of   \cite{renner2005information,   pirandola2009direct,
  garcia-patron2009continuous}, we  may ask whether the  AD statistics
Eq.  (\ref{eq:ADjp2})  can be  produced using only  local uncorrelated
classical noise  added by Alice  and Bob, starting from  the noiseless
case Eq. (\ref{eq:probs}). We now answer the question in the negative.

Alice's  most general  noise can  be modelled  by a  combination of  a
conditional probability distribution $P^A(x|y)$ (used with probability
$\alpha$) and  a random coin  toss $\varphi^A$ (used  with probability
$1-\alpha$),  while that  for Bob  by a  combination of  a conditional
probability  distribution $P^B(x|y)$  (used with  probability $\beta$)
and a random coin toss $\varphi^B$ (used with probability $1-\beta$).

Further, let  $P^A(0|0)=g$, $P^A(0|1)=h$ and  $P^B(0|0)=a, P^B(1|0)=b,
P^B(2|0)=c,   P^B(3|0)  =   1-a-b-c$;  and   $P^B(0|1)=d,  P^B(1|1)=e,
P^B(2|1)=f, P^B(3|1)  = 1-d-e-f$.  Applying the  noise unilaterally on
her side,  Alice can't reproduce  Eq. (\ref{eq:ADjp2}) because  of the
occurence of symbols 2 and 3 on Bob's side.  Suppose Bob alone applies
his  local  noise. Then,  one  finds  that  $P_{101}^B =  P_{110}^B  =
\frac{a+d}{8}$, which  stands in  contradiction with  the data  in Eq.
(\ref{eq:ADjp2}).  Thus, we  must consider whether both  Alice and Bob
applying  local   noise  independently  can  reproduce   the  required
statistics.  In the above, $P^A_j$  denotes the $j$th component of the
joint probability distribution obtained  by Alice's application of her
local   classical   noise   to   the   classical   outcome   data   of
Eq. (\ref{eq:probs}); analogously for $P^B_j$ in the case of Bob.

Without  loss of  generality, suppose  Alice applies  her local  noise
first,  and then  Bob.   We shall  use the  notation  where the  $j$th
component after Bob also has applied his local classical noisy channel
to  the  classical data  $P^A_j$  is  denoted $P^{A\rightarrow  B}_j$.
Then, from Eq. (\ref{eq:probs}), we obtain:
\begin{equation}
P_{010}^A = \frac{\alpha h}{8} + \frac{(1-\alpha)r}{8}.
\label{eq:or}
\end{equation}
This  must,  in  view  of  the vanishing  of  this  component  in  Eq.
(\ref{eq:ADjp2}), implying
\begin{subequations}
\begin{align}
\alpha=0,\quad r&= 0 \label{eq:cond1} \\
\alpha=1,\quad h&=0. \label{eq:cond2}
\end{align}
\label{eq:jodeq}
\end{subequations}
If $P^A_{010}$ doesn't  vanish, then we must have $p^B(0|0)  = 0$, to
ensure that under the transformation  induced by Bob's play, the final
$P^{A\rightarrow B}_{010}$ vanishes.  This  would mean that $p^B(1|0)$
or $p^B(2|0)$  or $p^B(3|0)$  should be  non-vanishing.  But  this, in
turn, would  mean that $P^{A\rightarrow B}_{011}$  or $P^{A\rightarrow
  B}_{012}$ or $P_{013}^{A\rightarrow B}$  should be non-vanishing, in
contradiction  with   the  corresponding   requirement  in   data  Eq.
(\ref{eq:ADjp2}).      Thus,    we     are    led     to    conditions
Eq. (\ref{eq:jodeq}).

To see why condition Eq. (\ref{eq:cond1}) won't work out, we note that
it   would  imply   that   $P^A_{000}   =  \frac{\alpha}{2}\left(g   +
\frac{h}{4}\right)  +  (1-\alpha)r\frac{5}{8}  \equiv 0$  as  well  as
$p^A_{001}  =   \frac{\alpha  h}{8}  +   (1-\alpha)\frac{r}{8}  \equiv
0$. But, this would imply that
\begin{align}
P^{A\rightarrow B}_{000} &= \beta(a P^A_{000} + d P^A_{001}) + 
(1-\beta)q(p^A_{000} + P^A_{001})  \nonumber\\
 &= 0,
\label{eq:000}
\end{align}
contradicting the fact that this  component is non-vanishing in the AD
statistics Eq. (\ref{eq:ADjp2}).
 
To see why condition Eq. (\ref{eq:cond2}) also won't work out, we note that
it   would  imply   that  
\begin{align}
P^{A\rightarrow B}_{000} &= \frac{g}{2}( \beta a + (1-\beta) q),\nonumber\\
P^{A\rightarrow B}_{100} &= \frac{g}{8}( \beta a + (1-\beta) q),
\label{eq:100}
\end{align}
implying that these two components differ by a factor 4, contradicting
the   additional  noise   dependence   seen  in  
Eq. (\ref{eq:ADjp2}). 

In conclusion, the advantage provided  by the quantum AD channel can't
be  simulated locally  (without  any classical  communication) by  the
legitimate parties, acting on the noiseless (but eavesdropped) outcome
statistics.   This  may be  attributed  to  the fundamentally  quantum
nature of  the disturbance introduced  into the noisy  channel through
Eve's intervention.

\section*{Acknowledgements} 
VS  thanks the  Ministery  of Human  Resource  Development, Govt.   of
India, for offering a doctoral  fellowship as a Ph.D. research scholar
at Indian Institute of Technology Jodhpur, Rajasthan, India. SB thanks
Atul  Kumar  and Anirban  Pathak  for  useful discussions  during  the
preliminary stage of this work. SB acknowledges support by the project
number  03(1369)/16/EMR-II   funded  by  Council  of   Scientific  and
Industrial Research, New Delhi, India. US and RS thank DST-SERB, Govt.
of  India,   for  financial  support  provided   through  the  project
EMR/2016/004019.
		
\bibliography{QKDnoise}

\end{document}